\documentclass{article}
\usepackage{spconf,amsmath,graphicx,amsfonts}
\usepackage[printonlyused,nolist]{acronym}
\usepackage[utf8]{inputenc}
\usepackage{stackengine,scalerel}
\usepackage{blindtext}
\usepackage{xcolor}

\begin{acronym}[SPACEEEEEE]
	\acro{SA}{simulated annealing}
	\acro{DL}{naive learning}
	\acro{ReLU}{rectified linear unit}
	\acro{MLP}{multilayer perceptron}
	\acro{NN}{neural network}
	\acro{SIF}{standard interference function}
	\acro{3GPP}{3rd generation partnership project}
	\acro{AP}{access point}
	\acro{AWGN}{additive white Gaussian noise}
	\acro{BF}{brute force}
	\acro{CSI}{channel state information}
	\acro{CSI}{channel state information}
	\acro{D2D}{device-to-device}
	\acro{FSPL}{free space path loss}
	\acro{GHz}{gigahertz}
	\acro{INR}{interference-to-noise ratio}
	\acro{ISI}{intersymbol interference}
	\acro{IoT}{Internet of Things}
	\acro{LoS}{line-of-sight}
	\acro{M2M}{machine-to-machine}
	\acro{MAC}{medium access control}
	\acro{MIMO}{multiple-input multiple-output}
	\acro{MTC}{machine-type communication}
	\acro{PSO}{particle swarm optimization}
	\acro{MT}{mobile terminal}
	\acro{MUSA}{multi-user shared access}
	\acro{NOMA}{non-orthogonal multiple access}
	\acro{NORA}{non-orthogonal resource allocation}
	\acro{OFDM}{orthogonal frequency division multiplexing}
	\acro{PHY}{physical layer}
	\acro{PL}{path loss}
	\acro{QoS}{quality-of-service}
	\acro{QuaDRiGa}{Quasi Deterministic Radio Channel Generator}
	\acro{RB}{resource block}
	\acro{RMS}{root mean square}
	\acro{RRM}{radio resource management}
	\acro{SER}{symbol-error rate}
	\acro{SINR}{signal-to-interference-plus-noise ratio}
	\acro{SIR}{signal-to-interference ratio}
	\acro{SNR}{signal-to-noise ratio}
	\acro{UE}{user equipment}
	\acro{cMTC}{critical machine-type communications}
	\acro{dB}{decibel}
	\acro{eMBB}{enhanced mobile broad-band}
	\acro{mmWave}{millimeter-wave}
	\acro{nLoS}{non-line-of-sight}
	\acro{SA}{simulated annealing}
	\acro{FP}{fixed point}
	\acro{CF}{cost function}
	\acro{ML}{machine learning}
	\acro{DNN}{deep neural network}
	\acro{UMi}{urban microcell}
\end{acronym}

\makeatletter
\def\SavedSubStyle{\csname @mstyle\s@switch\endcsname}
\renewcommand\ThisStyle[1]{%
\ifmmode%
\def\@mmode{T}\mathchoice%
{\edef\m@switch{D}\edef\s@switch{S}\LMex=1ex\relax\LMpt=1pt\relax#1}%
{\edef\m@switch{T}\edef\s@switch{S}\LMex=1ex\relax\LMpt=1pt\relax#1}%
{\edef\m@switch{S}\edef\s@switch{s}\LMex=\scriptstyleScaleFactor ex\relax%
\LMpt=\scriptstyleScaleFactor pt\relax#1}%
{\edef\m@switch{s}\edef\s@switch{s}\LMex=\scriptscriptstyleScaleFactor ex\relax%
\LMpt=\scriptscriptstyleScaleFactor pt\relax#1}%
\else%
\def\@mmode{F}%
\edef\m@switch{T}\edef\s@switch{S}\LMex=1ex\relax\LMpt=1pt\relax#1%
\fi%
}
\makeatother

\newcommand\overstar[1]{\ThisStyle{\ensurestackMath{%
  \setbox0=\hbox{$\SavedStyle#1$}%
  \stackengine{1.0pt}{\copy0}{\kern.2\ht0\smash{\SavedSubStyle\star}}{O}{c}{F}{T}{S}}}}

\newcommand{\overbar}[1]{\mkern 2.6mu\overline{\mkern-1.5mu#1\mkern-1.5mu}\mkern 1.5mu}
\def\setN{\mathcal{N}}
\def\setM{\mathcal{M}}
\def\pwrbdgt{\overbar{P}}

\def\rxbw{\boldsymbol{\theta}^{\text{Rx}}}
\def\rxbwbar{\overbar{\boldsymbol{\theta}}^{\text{Rx}}}
\def\rxbd{\boldsymbol{\beta}^{\text{Rx}}}
\def\rxbdbar{\overbar{\boldsymbol{\beta}}^{\text{Rx}}}
\def\rxbwopt{\overstar{\boldsymbol{\theta}}^{\text{Rx}}} 
\def\rxbdopt{\overstar{\boldsymbol{\beta}}^{\text{Rx}}}
\def\realN{\mathbb{R}^N}
\def\realM{\mathbb{R}^M}
\def\disMtheta{\mathbb{D}^M_{\theta}}
\def\disMbeta{\mathbb{D}^M_{\beta}}
\def\distheta{\mathbb{D}_{\theta}}
\def\disbeta{\mathbb{D}_{\beta}}
\def\ueN{\ac{UE}\hspace{0.6pt}$_n$\hspace{1pt}}

\newcommand{\ma}[1]{{\mathbf{#1}}}
\newcommand{\maxx}[1]{{\underset{#1}{\text{max}}}}
\newcommand{\amin}[1]{{\underset{#1}{\text{arg min}}}}
\newcommand{\amax}[1]{{\underset{#1}{\text{arg max}}}}
\newcommand{\nbrace}[1]{{\left ( {#1} \right )}}
\newcommand{\cbrace}[1]{{\left \{ {#1} \right \}}}
\newcommand{\sbrace}[1]{{\left [ {#1} \right ]}} 
\newcommand{\norm}[1]{{\left \| {#1} \right \|}}
\newcommand{\cardinality}[1]{{\left | {#1} \right |}}
\newcommand{\fu}[1]{{\mathcal{#1}}}
\newcommand{\veb}[1]{{\boldsymbol{#1}}}

\title{Deep Learning Beam Optimization in \\Millimeter-Wave Communication Systems}
\name{Rafail Ismayilov$\hspace{1pt}^*$, Renato L. G. Cavalcante$\hspace{1pt}^{* \dag}$ and Sławomir Stańczak$\hspace{1pt}^{* \dag}$}
\address{$^*$Fraunhofer Heinrich Hertz Institute, Berlin, Germany\\	
	$^\dag$Technical University of Berlin, Berlin, Germany \\}

\begin{document}
\frenchspacing
\ninept
\interfootnotelinepenalty=10000
\maketitle
\begin{abstract}
We propose a method that combines fixed point algorithms with a neural network to optimize jointly discrete and continuous variables in millimeter-wave communication systems, so that the users' rates are allocated fairly in a well-defined sense. In more detail, the discrete variables include user-access point assignments and the beam configurations, while the continuous variables refer to the power allocation. The beam configuration is predicted from user-related information using a neural network. Given the predicted beam configuration, a fixed point algorithm allocates power and assigns users to access points so that the users achieve the maximum fraction of their interference-free rates. The proposed method predicts the beam configuration in a ``one-shot'' manner, which significantly reduces the complexity of the beam search procedure. Moreover, even if the predicted beam configurations are not optimal, the fixed point algorithm still provides the optimal power allocation and user-access point assignments for the given beam configuration.
\end{abstract}
\begin{keywords}
	Interference management, millimeter-wave communication, beamforming, deep learning, fixed point algorithm
\end{keywords}
\acresetall 
\vspace{-4pt}
\section{Introduction}
\vspace{-4pt}
\label{sec:intro}
\acresetall 
Communication over \ac{mmWave} bands is a key enabler to support increasing data rate demands \cite{niu2015survey}.
Small wavelengths allow the exploitation of large antenna arrays in the current size of radio chips, which results in a substantial gain in the link budget using beamforming.
Such a gain can largely compensate for the high path-loss in the \ac{mmWave} band, without increasing the transmit power. 
Achieving higher gain, however, requires narrow beams both at the \ac{UE} and at the \ac{AP}. 
The latter, in turn, needs to deal with difficult problems associated with the establishment and maintenance of a robust communication links with directional beams. 
Thus, resource management in \ac{mmWave} systems becomes more difficult owing to additional system parameters such as beam width and beam direction, in comparison with conventional communication technologies with the omni- or semi-directional transmission.  
More specifically, in addition to the power allocation and \ac{UE}-\ac{AP} assignment problems in conventional communication systems, the network operating with directional beams further requires to find the optimal beam configurations to improve system performance with respect to network coverage, spectral efficiency, etc.

In practical \ac{mmWave} communication systems, some transceivers are implemented with discrete beam configurations selected from a predefined finite set \cite{beamforming_standard}.
As a result, joint optimization of all discrete and continuous variables becomes difficult.
For example, existing methods in the literature for power allocation \cite{Zeng2019, kwon2018multi}, beam configuration optimization \cite{Liu_2020, Ma_2021,Ismayilov2018} and \ac{UE}-\ac{AP} assignment \cite{Aboagye_2019, Chatterjee_2020} cope with some of the continuous or the discrete variables, but the joint optimization of all these parameters is difficult and requires heuristics.
For example, in a previous study \cite{Ismayilov2019}, a method that jointly optimizes the transmit power of \acp{UE}, receive beam configurations of \acp{AP}, and \ac{UE}-\ac{AP} assignments has been proposed.
In that method, for fixed beam configurations of \acp{AP}, the problem is formulated as a weighted rate allocation problem, where each \ac{UE} maximizes the same portion of its maximum achievable rate that it would have in interference-free conditions.
The solution to the aforementioned problem is obtained with an iterative fixed point algorithm that allocates power and assigns \acp{UE} to \acp{AP}. 
However, the algorithm is unable to include the beam configurations in the optimization framework while guaranteeing optimality. 
As a result, the study in \cite{Ismayilov2019} combines fixed point algorithms with heuristics based on \ac{SA} to cope with all optimization variables.
Despite the high performance of the method in \cite{Ismayilov2019}, from a practicality perspective, one of the challenges is to alleviate the complexity of the heuristic based on simulated annealing, which scales exponentially with the number of discrete beam configurations and the number of \acp{AP}.

In this work, we propose a method that combines fixed point algorithms with a neural network, with the intent of jointly optimizing all discrete and continuous variables.
To this end, we propose an architecture based on a deep neural network that is able to predict the beam configurations from \ac{UE} related information.
More precisely, we use a neural network for supervised multitask learning where each \ac{AP} learns its best beam configuration in the sense of maximizing the common fraction of the maximum achievable rates of the \acp{UE}.
Given the predicted beam configurations, the fixed point algorithm provides the solution for optimal transmit power allocation and the \ac{UE}-\ac{AP} assignments.
With such a combination of a neural network with a fixed point algorithm, we have the following advantages.
First, compared to the iterative \ac{SA}, the neural network can predict the beam configuration in a ``one-shot" manner, which results in a significant complexity reduction.
Second, if the beam configurations produced by the neural network are suboptimal, then the fixed point algorithm is still optimal in the sense of maximizing the common fraction of interference-free rates for the given beam configurations.
Third, with the settings considered here, the neural network can produce robust predictions under environmental changes (e.g., different distributions of \ac{UE} positions).
\section{System Model and Problem Statement}
\label{sec:format}
\subsection{System Model}
\label{system_model}
In this study, we use the following standard notations.
By $\norm{\cdot}_{\infty }$, we denote the standard $l - \infty$ norm.
Sets of non-negative and strictly positive reals are denoted by $\mathbb{R}_{+}$ and $\mathbb{R}_{++}$, respectively.

We consider a wireless network with a set $\fu{N} = \cbrace{1,...,N}$ of transmitters, called \acp{UE}, and a set $\fu{M} = \cbrace{1,...,M}$ of receivers, called \acp{AP}.
The \ac{UE} transmit beam configurations are denoted by $\nbrace{\forall n \in \mathcal{N}} \hspace{2pt} \theta^{\text{Tx}}_n$ and $\beta^{\text{Tx}}_n$, where $\theta^{\text{Tx}}_n$ and $\beta^{\text{Tx}}_n$ are the $\text{UE}_n$ transmit beam width and direction, respectively.
We assume that the transmit beam widths of \acp{UE} are identical and fixed $\nbrace{\forall n \in \mathcal{N}} \hspace{2pt} \theta_n^{\text{Tx}}=\overbar{\theta}^{\text{Tx}}$, while \ac{UE} beam directions are uniformly distributed with $\nbrace{\forall n \in \mathcal{N}} \hspace{2pt}\beta_n^{\text{Tx}} \sim \mathcal{U}\nbrace{\beta^{\text{Tx}}_{\text{min}},\beta^{\text{Tx}}_{\text{max}}}$.
In contrast to the \acp{UE}, each \ac{AP} beam width and direction can be adjusted by the respective \ac{AP}, i.e. $\rxbw = \nbrace{\theta_1^{\text{Rx}},...,\theta_M^{\text{Rx}}} \in \disMtheta \subseteq \realM_{++}$ and $\rxbd = \nbrace{\beta_1^{\text{Rx}},...,\beta_M^{\text{Rx}}} \in \disMbeta \subseteq \realM$ are vectors to be optimized.
Note that $\distheta$ and $\disbeta$ denote the discrete sets of receive beam widths and directions, respectively.
The transmit power vector of \acp{UE} is defined by $\veb{p} := \nbrace{p_1,...,p_N} \in \realN_+$, and the elements of this vector take values in the interval $\nbrace{\forall n \in \mathcal{N}} \hspace{2pt}  0 < p_n \le \pwrbdgt$, where $\pwrbdgt$ is the maximum allowed transmit power.
In this work, we assume that multiple \acp{UE} may simultaneously connect to an \ac{AP}.

For given $\veb{p},\rxbw$, and $\rxbd$, the achievable rate of \ueN in the uplink to its best serving \ac{AP} (i.e. the \ac{UE}-\ac{AP} assignment) is expressed by
\begin{equation}\label{eq.achievable_rate}
	\resizebox{0.91\hsize}{!}{$
		R_n \nbrace{\veb{p},\rxbw,\rxbd} = \maxx{m \in \fu{M}} W \log_2 \Big( 1 + s_n \nbrace{\veb{p},\rxbw,\rxbd,m} \Big),
	$}
\end{equation}
\noindent where $W$ is the system bandwidth, and 
\begin{equation}\label{eq.SINR}
\resizebox{0.91\hsize}{!}{$
\begin{array}{rcl}
s_n: \realN_+ \times \realM_{++} \times \realM \times \mathcal{M} & \rightarrow  & \mathbb{R}_+ \\
\nbrace{\veb{p}, \rxbw,\rxbd, m} & \mapsto & \dfrac{p_n h_{m,n} \nbrace{\rxbw,\rxbd}}{ \displaystyle \sum_{n' \in  \mathcal{N}  \setminus \cbrace{n} } p_{n'} h_{m,n'}\nbrace{\rxbw,\rxbd} + \sigma^2}
\end{array}
$}
\end{equation}

\noindent denotes the \ac{SINR} for UE$_n$.
The variable $\sigma^2 > 0$ is the noise power, and $\nbrace{\forall m \in \setM}$ $\nbrace{\forall n \in \setN}$ $\nbrace{\forall \nbrace{\rxbw,\rxbd} \in \mathcal{Q} }$ $ h_{m,n}\nbrace{\rxbw,\rxbd} > 0$ denotes the channel power gain between \ac{UE}$_n$ and serving \ac{AP}$_m$ given beam configuration $\nbrace{\rxbw,\rxbd} \in \mathcal{Q}$, as defined in \cite[Sect.~II.C]{Ismayilov2019}.
By fixing  $\nbrace{\forall n \in \mathcal{N}} \hspace{2pt}  p_n = \pwrbdgt$, the maximum achievable rate, also called interference-free rate, of \ueN is given by
\begin{equation}
\overbar{R}_n = \maxx{m \in \fu{M}, \nbrace{\rxbw,\rxbd} \in \mathcal{Q}} W \log_2 \nbrace{1 + \dfrac{\pwrbdgt h_{m,n}\nbrace{\rxbw,\rxbd}}{\sigma^2}}.
\end{equation}
The rate $\overbar{R}_n$ corresponds to the case when \ueN transmits alone in the network with full power to its best serving \ac{AP}.
\subsection{Problem Statement}
The problem in this study is defined as fair allocation of the \ac{UE} rates $\nbrace{\forall n \in \fu{N}} \hspace{2pt} R_n$, in the sense that every \ac{UE} achieves the maximum common fraction $c \in \sbrace{0,1}$ of its interference-free rates $\overbar{R}_n$.
Formally, the optimization problem is stated as follows:
\begin{subequations} \label{problem_statement}
	\begin{alignat}{3}
		\underset{\veb{p},\rxbw,\rxbd, c}{\text{maximize}} & \quad c                                                                                    & \quad  \tag{\ref{problem_statement}}  \\  
		\text{subject to}                                  & \quad \nbrace{ \forall n \in \fu{N}}\hspace{2pt} c \overbar{R}_n = R_n \nbrace{\veb{p},\rxbw,\rxbd}    & \label{problem_statement_a}         \\
		                                                   & \quad \norm{\veb{p}}_{\infty } \le \overbar{P}                                              & \label{problem_statement_b}         \\
		                                                   & \quad \veb{p} \in \realN_+, \nbrace{\rxbw,\rxbd} \in \fu{Q}, c \in \mathbb{R}_{++},         & \label{problem_statement_c}         
	\end{alignat}
\end{subequations}

\noindent where $\fu{Q} = \cbrace{\nbrace{\rxbw_k,\rxbd_l}}_{k \in \nbrace{1,...,\cardinality{\distheta}^M}, \hspace{2pt} l \in \nbrace{1,...,\cardinality{\disbeta}^M}}$ is a set of all receive beam configurations of the \acp{AP}.
The joint optimization of the parameters in \eqref{problem_statement} is challenging due to a mixture of discrete and continuous parameters.
\begin{figure*}[htb]
	\centering
	\centerline{\includegraphics[width=0.91\textwidth]{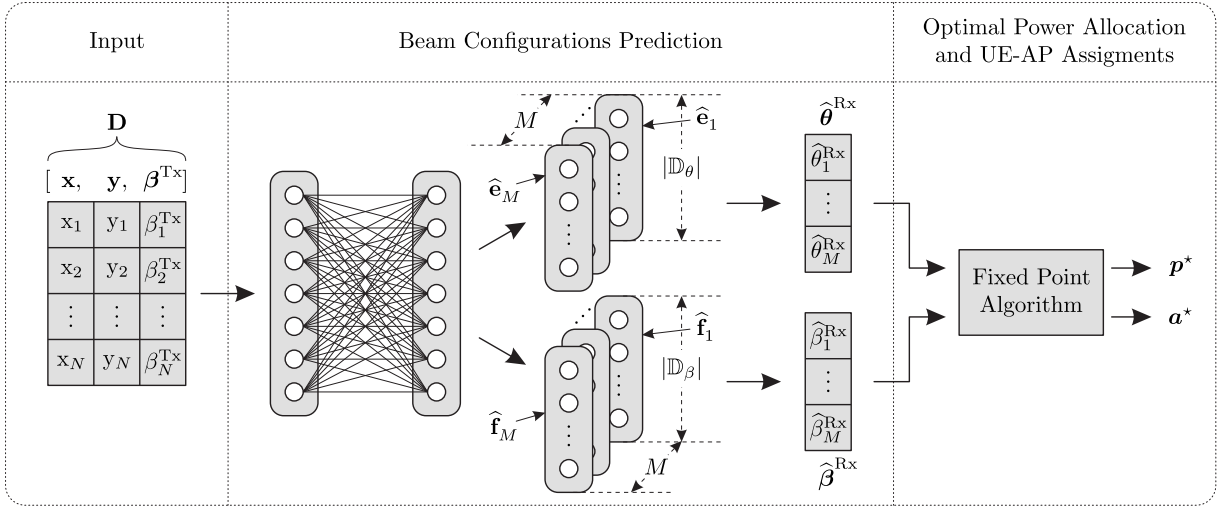}}
	\vspace{-1pt}
	\caption{The proposed architecture for joint optimization of the beam configurations, power allocation and \ac{UE}-\ac{AP} assignments. The neural network predicts the beam configurations from \ac{UE} related information $\ma{D}$, and the fixed point algorithm optimally allocates the power and assigns the \acp{UE} to \acp{AP} given the predicted beam configurations.}
	\label{neural_network_diagram}
\end{figure*}
However, if the tuple $\nbrace{\rxbw,\rxbd}$ is fixed to any given beam configuration $\nbrace{\rxbwbar,\rxbdbar} \in \fu{Q}$, the objective reduces to the following power allocation and \ac{UE}-\ac{AP} assignment problem:
\begin{subequations} \label{reduced_problem}
	\begin{alignat}{3}
		\underset{\veb{p}, c}{\text{maximize}} & \quad c                                                                                     & \quad \tag{\ref{reduced_problem}} \\
		\text{subject to}                                  & \quad \nbrace{ \forall n \in \fu{N}}\hspace{2pt} c \overbar{R}_n = R_n \nbrace{\veb{p},\rxbwbar,\rxbdbar}, & \label{reduced_problem_a}         \\
		                                                   & \quad \norm{\veb{p}}_{\infty } \le \overbar{P}                                              & \label{reduced_problem_b}         \\
		                                                   & \quad \veb{p} \in \realN_+, c \in \mathbb{R}_{++},        &  \label{reduced_problem_c} 
	\end{alignat}
\end{subequations}

\noindent and problem \eqref{reduced_problem} can be solved optimally with a simple iterative fixed point algorithm \cite{Cavalcante_2019}\cite[Sect.~IV.A]{Ismayilov2019}.

The connection between \eqref{problem_statement} and \eqref{reduced_problem} can be summarized as follows. 
Suppose that $\nbrace{\rxbwopt, \rxbdopt} \in \fu{Q}$ is an optimal beam configuration to problem \eqref{problem_statement}.
If we solve \eqref{reduced_problem} by fixing $\rxbwbar = \rxbwopt$ and $\rxbdbar = \rxbdopt$, then the solution $\nbrace{c^{\star},\veb{p}^{\star}}$ to \eqref{reduced_problem} is also the optimal fraction $c^{\star}$ and an optimal power allocation $\veb{p}^{\star}$ to problem \eqref{problem_statement}. 
Furthermore, the entries of the optimal \ac{UE}-\ac{AP} assignment vector $\veb{a}^{\star} = \nbrace{a_1^{\star},..., a_N^{\star}} \in \mathcal{M} \times \cdots \times \mathcal{M}$ can be recovered via
\begin{equation}
a^{\star}_n \in \amin{ m \in \setM} \dfrac{\overbar{R}_n p^{\star}_n}{\vspace{10pt}W \log_2 \Big( 1 + s_n \nbrace{\veb{p}^{\star},\rxbwopt,\rxbdopt,m} \Big)}.
\end{equation}

Therefore, if the optimal beam configurations are known, \eqref{problem_statement} can be solved optimally with a simple fixed point algorithm.
In principle, the optimal beam configurations could be found via exhaustive search or other iterative search methods (e.g. \ac{SA}) over $\mathcal{Q}$, as done in \cite{Ismayilov2019}.
Performing this search, however, scales exponentially with the number of discrete beam configurations and the number of \acp{AP}.
In the following, we propose a deep learning-based non-iterative method that predicts the beam configurations in a ``one-shot'' manner from \ac{UE} related information.

\section{Beam Configurations Prediction with Deep Neural Network}
To find the optimal beam configuration $\nbrace{\rxbwopt,\rxbdopt}$ in \eqref{problem_statement}, we assume that there exists an ideal mapping $\Xi$ that maps the \ac{UE} related information $\ma{D}$ to some output $\big( \sbrace{\ma{e}_1^{\star},..,\ma{e}_M^{\star}}, \sbrace{\ma{f}_1^{\star},..,\ma{f}_M^{\star}} \big)$ (to be explained later) such that $\rxbwopt$ and $\rxbdopt$ can be reconstructed from $\sbrace{\ma{e}_1^{\star},..,\ma{e}_M^{\star}}$ and $\sbrace{\ma{f}_1^{\star},..,\ma{f}_M^{\star}}$, respectively, and the reconstructed configuration $\nbrace{\rxbwopt,\rxbdopt}$ is the optimal beam configuration to problem \eqref{problem_statement}.
Formally, the ideal mapping $\Xi$ is defined as follows
\begin{equation}
\resizebox{0.91\hsize}{!}{$
\Xi : \mathbb{R}^{N \times 3} \rightarrow \mathbb{R}^{M \times |\distheta|} \times \mathbb{R}^{M \times |\disbeta|} : \ma{D} \mapsto \big( \sbrace{\ma{e}_1^{\star},..,\ma{e}_M^{\star}}, \sbrace{\ma{f}_1^{\star},..,\ma{f}_M^{\star}} \big),
$}
\end{equation}
where $\ma{D} = \sbrace{\ma{x}, \ma{y}, \boldsymbol{\beta}^{\text{Tx}}}$ is the matrix that contains \ac{UE} location information in a 2D Cartesian coordinate system $\sbrace{\ma{x}, \ma{y}}$, and $\boldsymbol{\beta}^{\text{Tx}}$ is the \ac{UE} beam direction.
The rows of the matrix $\ma{D}$ are sorted in lexicographical order.

We now proceed to explain the output of the ideal mapping $\Xi$, and we also describe the reconstruction mechanism to obtain $\rxbwopt$ and $\rxbdopt$ from $\sbrace{\ma{e}_1^{\star},..,\ma{e}_M^{\star}}$ and $\sbrace{\ma{f}_1^{\star},..,\ma{f}_M^{\star}}$.
We assume that the elements of the vectors $\nbrace{\forall m \in \setM} \hspace{2pt} \ma{e}^{\star}_m, \ma{f}^{\star}_m$ are one-hot encoded, i.e., $\nbrace{\forall m \in \setM} \hspace{2pt} \ma{e}^{\star}_m \in \cbrace{0,1}^{|\distheta|}, \ma{f}^{\star}_m \in \cbrace{0,1}^{|\disbeta|}$.
In this way, the indices of the vectors $\nbrace{\forall m \in \setM} \hspace{2pt} \ma{e}_m^{\star}, \ma{f}_m^{\star}$ are associated with the indices of the respective beam width and direction indices from the discrete sets
$\distheta$ and $\disbeta$.
To relate the selected optimal beam indices (in the sense of maximizing the objective in \eqref{problem_statement}), we further assume that the vectors $\nbrace{\forall m \in \setM} \hspace{2pt} \ma{e}_m^{\star}, \ma{f}_m^{\star}$ contain the value one at the index corresponding to the selected optimal beam width and direction, respectively, and zeros elsewhere.

With the output explained above, the reconstruction of the beam configurations $\rxbwopt$ and $\rxbdopt$ is performed as follows.
Once we obtain the output $\big( \sbrace{\ma{e}_1^{\star},..,\ma{e}_M^{\star}}, \sbrace{\ma{f}_1^{\star},..,\ma{f}_M^{\star}} \big)$ from $\Xi$, the vectors $\rxbwopt$ and $\rxbdopt$ are recovered via
\begin{equation}\label{rule_theta}
\rxbwopt = \nbrace{\overstar{\theta}^{\text{Rx}}_1,...,\overstar{\theta}^{\text{Rx}}_M}, \nbrace{\forall m \in \setM} \begin{array}{l}
\overstar{\theta}_m^{\text{Rx}} = \gamma_{i^{\star}_m} \in \distheta\\ 
i^{\star}_m \in \amax{i = 1,...,|\distheta|} \nbrace{ \sbrace{\ma{e}^{\star}_m}_i }
\end{array}
\end{equation}
and
\begin{equation}\label{rule_beta}
\resizebox{0.91\hsize}{!}{$
\rxbdopt = \nbrace{\overstar{\beta}^{\text{Rx}}_1,...,\overstar{\beta}^{\text{Rx}}_M}, \nbrace{\forall m \in \setM} \begin{array}{l}
\overstar{\beta}_m^{\text{Rx}} = \delta_{i^{\star}_m} \in \disbeta\\ 
i^{\star}_m \in \amax{i = 1,...,|\disbeta|} \nbrace{ \sbrace{\ma{f}^{\star}_m}_i } ,
\end{array}
$}
\end{equation}
where $\gamma_{i^{\star}_m}$ and $\delta_{i^{\star}_m}$ represent the selected optimal beam configurations from the finite and discrete sets $\distheta$ and $\disbeta$, respectively.
Although the ideal mapping $\Xi$ provides optimal beam configurations by assumption, it is challenging to analytically characterize it. Therefore we propose a deep neural network that learns an ideal mapping $\Xi$ from data.

The proposed neural network architecture in combination with the fixed point algorithm is shown in Fig.~\ref{neural_network_diagram}.
With the setting that the beam configurations are taken from the discrete and finite sets $\distheta$ and $\disbeta$, we consider the beam configurations as labels, and we pose the beam optimization problem in \eqref{problem_statement} as a multi-label classification problem \cite{Tsoumakas2007}.
The labels are constructed as follows.
Given the finite set $\mathcal{Q}$ consisting of all possible beam configurations, and the \ac{UE} related information $\ma{D}$, we perform an exhaustive search by applying fixed point iterations for every candidate beam configuration $\nbrace{\rxbw,\rxbd} \in \fu{Q}$ to obtain $\nbrace{\rxbwopt, \rxbdopt}$.
\footnote{If exhaustive search is too complex to construct the training set, the approach can be easily adapted to use heuristics such as simulated annealing \cite{Ismayilov2019}. The construction of training sets can be done offline with time-consuming heuristics because there is no real-time communication taking place.}
Next, we retrieve $\nbrace{m \in \mathcal{M}} \hspace{2pt} i^{\star}_m$ by following the rules in \eqref{rule_theta} and \eqref{rule_beta}.
With the optimal index $i^{\star}_m$ we construct the one-hot encoded ground truth labels $\big( \sbrace{\ma{e}_1^{\star},..,\ma{e}_M^{\star}}, \sbrace{\ma{f}_1^{\star},..,\ma{f}_M^{\star}} \big)$.

Given the input-output pair $\Big( \ma{D}, \big( \sbrace{\ma{e}_1^{\star},..,\ma{e}_M^{\star}}, \sbrace{\ma{f}_1^{\star},..,\ma{f}_M^{\star}} \big) \Big)$, we extract the features from $\ma{D}$ by utilizing two fully connected layers with the \ac{ReLU} activation as illustrated in Fig.~\ref{neural_network_diagram}.
To project the extracted features onto beam configuration labels, we customize the multi-label classification layer with the softmax activation function so that $\nbrace{ \forall m \in \setM } \hspace{2pt} \widehat{\ma{e}}_m \in \sbrace{0,1}^{|\distheta|}$ and $\widehat{\ma{f}}_m \in \sbrace{0,1}^{|\disbeta|}$.
With this design, the output of the neural network indicates the beam configuration indices.
In other words, the indices of the highest values in $\nbrace{ \forall m \in \setM } \hspace{2pt} \widehat{\ma{e}}_m, \widehat{\ma{f}}_m$ indicates the best predicted beam width and direction for AP$_m$.
We train the neural network by trying to minimize the categorical cross entropy loss \cite{Mannor_cros_entropy} given by
\begin{equation}
\resizebox{0.90\hsize}{!}{$
\begin{array}{c}
l \nbrace{ [\ma{e}_1^{\star},..,\ma{e}_M^{\star}], [\ma{f}_1^{\star},..,\ma{f}_M^{\star}], [\widehat{\ma{e}}_1,..,\widehat{\ma{e}}_M], [\widehat{\ma{f}}_1,..,\widehat{\ma{f}}_M]}  =  \vspace{5pt}\\

= - \displaystyle\sum_{m=1}^{M} \sbrace{ \displaystyle\sum_{i=1}^{|\distheta|} \sbrace{ \ma{e}^{\star}_m }_i \log \nbrace{\sbrace{ \widehat{\ma{e}}_m }_i} + \displaystyle\sum_{i=1}^{|\disbeta|} \sbrace{ \ma{f}^{\star}_m }_i \log ( [ \widehat{\ma{f}}_m ]_i)}.
\end{array}
$}
\end{equation}

\section{Numerical Results}
\subsection{Training Parameters and Reference Methods}
To train the proposed neural network, we generate $10^6$ samples, and we split those samples into training and test sets containing $9 \cdot 10^5$ and $1 \cdot 10^5$ samples, respectively.
The samples are generated using the mmWave channel model described in \cite{5GCM_2016}.
The channel model considers an \ac{UMi}-\ac{LoS} scenario with high user density in open areas and street canyons.
The neural network has two hidden layers, each consisting of 200 fully connected neurons.
The total number of epochs and batch size for training is 500 and 512, respectively.
We train a neural network using the \emph{Adadelta} optimizer \cite{zeiler2012adadelta} with adaptive learning rates.

We compare the performance of the proposed method with exhaustive search and with the method in \cite{Ismayilov2019}.
In addition, we also introduce a method called naive learning. 
It simply provides to the fixed point algorithm the beam configurations encountered most frequently in the training set.
\subsection{Simulation}
\begin{figure}[ht]	
		\centering
		\centerline{\includegraphics[width=80.0mm]{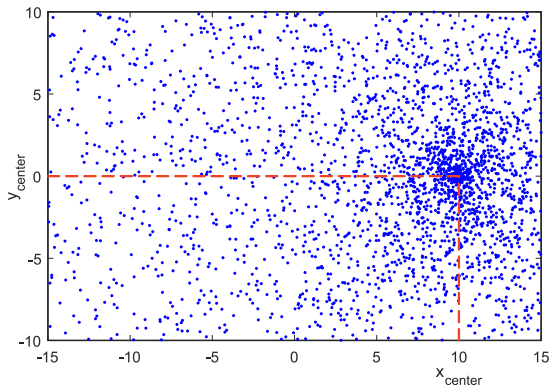}}
		\vspace{-10pt}
	\caption{Sample scatter plot of \ac{UE} positions in the test set.}
	\label{illustration_of_enviromental_change}
\end{figure}

\vspace{6pt}
Performing an exhaustive search over a large solution set $\mathcal{Q}$ is infeasible, so, for the simulations, we consider a small scale problem with $N=10$ \acp{UE}, $M=3$ \acp{AP}, discrete receive beam width configurations $\distheta=\cbrace{30^{\circ},45^{\circ},60^{\circ}}$ and discrete receive beam direction configurations $\disbeta=\cbrace{80^{\circ},90^{\circ},100^{\circ}}$.
The \acp{UE} are located in a $20 \times 30$ m$^2$ rectangle area centered at $\sbrace{0\text{m},0\text{m}}$.
With the given parameters, the size of the solution set is $\cardinality{\fu{Q}} = \cardinality{\distheta}^M \times \cardinality{\disbeta}^M = 729$.
We set the number of fixed point iterations to 100.
The heuristic based on simulated annealing proposed in \cite{Ismayilov2019} selects the initial beam configurations from the solution set $\fu{Q}$ uniformly at random, without prior knowledge about the configuration of the \ac{mmWave} network.
We formally define the optimal rate fraction $c^{\star}$ obtained by exhaustive as 100 \% \emph{solution efficiency}, and we compare the relative performances of other methods with the optimal solution obtained by exhaustive search.
To evaluate the robustness of the proposed neural network against distribution changes in the input data, we consider the following two cases:
\vspace{7pt}
\begin{itemize}
\item[\textbf{C1:}] The \ac{UE} positions $\sbrace{\ma{x},\ma{y}}$ in $\ma{D}$ are selected randomly from a uniform distribution within the rectangle area for both training and testing. 
\item[\textbf{C2:}] The \ac{UE} positions $\sbrace{\ma{x},\ma{y}}$ in $\ma{D}$ are selected randomly from a uniform distribution within the rectangle area only for the training set.
The \ac{UE} positions $\sbrace{\ma{x},\ma{y}}$ in the test set is selected randomly from a non-uniform distribution within the rectangle area.
More precisely, we consider two uniformly distributed random variables $r \in [0, L]$ and $\phi \in [0, 2 \pi]$, where $L$ denotes the radius of a disk, and we generate the \ac{UE} coordinates $[\text{x},\text{y}]$ in 2D Cartesian coordinate system as
\begin{equation}
\text{x} = \text{x}_{\text{center}} + r \cos \phi \hspace{5pt} \text{and} \hspace{5pt} \text{y} = \text{y}_{\text{center}} + r \sin \phi ,
\end{equation}
where $\text{x}_{\text{center}}$ and $\text{y}_{\text{center}}$ denote the center of a disk on $X$ and $Y$ axis, respectively.
A scatter plot obtained with this distribution is illustrated in Fig.~\ref{illustration_of_enviromental_change}.
Note that only the \ac{UE} positions within the rectangle area were used for the simulation as shown in Fig.~\ref{illustration_of_enviromental_change}.
\end{itemize}

\begin{figure}[t]
	\centering
	\centerline{\includegraphics[width=8.6cm]{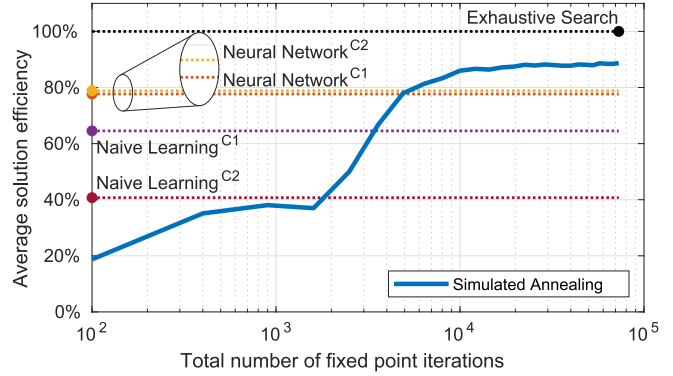}}
		\vspace{-10pt}
	\caption{Average solution efficiency as a function of the number of fixed point iterations.}
	\label{fig3}
	\vspace{-8pt}
\end{figure}

The performance of the proposed and reference methods are given in Fig.~\ref{fig3}.
The superscripts $^\text{C1}$ and $^\text{C2}$ in Fig.~\ref{fig3} denote the performance of the neural network and the naive learning approaches according to the previously-mentioned cases.
Since all methods other than the simulated annealing and the exhaustive search method are predicting the beam configurations in a ``one-shot'' manner, we mark their performance with points, and we extend the dashed lines with the respective \emph{solution efficiency} values in the X-axis.

In Fig.~\ref{fig3}, we notice a substantial gap between the performance of the neural network and the naive learning approach under \textbf{C1}.
This result shows that the neural network is able to predict good beam configurations from \ac{UE} related information. 
The second, and very attractive result in Fig.~\ref{fig3}, is the similar performance of the neural network under \textbf{C2} and \textbf{C1}. 
This result indicates that the distribution used for training is likely to provide good performance when the distribution of the test set differs.
In addition, we note that the proposed scheme requires only a single call to a fixed point algorithm (100 iterations are used) to achieve 80 \% of the optimal rates on average. 
In contrast, the approach in \cite{Ismayilov2019} and exaustive search require one call to the fixed point algorithm for each configuration being probed. 
As a result, the number of iterations of the fixed point algorithm required by those heuristics are orders of magnitude larger than that required by the propose scheme to achieve the same performance. This result indicates that the proposed scheme can be useful for real-time operation. 


\vspace{-6pt}
\section{Conclusion}
In this work, we proposed a method that jointly optimizes the beam configuration, \ac{UE}-\ac{AP} assignments, and the power allocation in \ac{mmWave} communication systems with directional transmission.
We introduced a neural network that predicts the beam configuration from \ac{UE} related information, and the fixed point algorithm solves the \ac{UE}-\ac{AP} assignment and power allocation problems given the predicted beam configuration.
One of the advantages of the proposed method is that it predicts the beam configuration non-iteratively, which is an important factor in \ac{mmWave} communication systems with low-latency requirements.
Another advantage of the proposed method is that the fixed point algorithm guarantees the optimal power allocation and \ac{UE}-\ac{AP} assignments given any prediction from the neural network.
Simulations showed that the proposed method can provide 80 \% of the optimal rates on average with a single call of the fixed point algorithm, while the reference method based on simulated annealing requires to solve fixed point problems repeatedly to achieve the same performance.
\clearpage
\bibliographystyle{IEEEbib}
\bibliography{strings,refs}

\end{document}